\documentstyle[preprint,aps,epsf]{revtex}

\def\Lam{\Lambda}  \def\Sig{\Sigma}
\def\wave{\simeq}
\def\ie{{\it i.e.}}
\def\di{{\Delta I}}
\def\pip{{\pi^+}}  \def\pim{{\pi^-}}

\title{${\pi^+}$ Emission from Hypernuclei and the 
Weak ${\di=3/2}$ Transitions}
\author{Makoto Oka%
        \thanks{email: oka@th.phys.titech.ac.jp}}
\address{Department of Physics, Tokyo Institute of Technology \\
        Meguro, Tokyo, 152-8551 JAPAN}

\date{\today}

\begin{document}

\maketitle

\begin{abstract}
Low energy $\pip$ emission in hypernuclear weak decays is studied in 
the soft pion limit.  It is found that the $\pip$ decay amplitude is 
dominated by the $\di=3/2$ part of nonmesonic weak decays according to the 
soft pion theorem.  
The ratios, $R(\pip/\hbox{soft $\pim$})$, for light hypernuclei are 
estimated in the direct quark mechanism, which predicts violation of 
the $\di=1/2$ rule, and found to be about $1/3$ for $^4_{\Lam}He$. 
\end{abstract}

\newpage

Recent experimental and theoretical studies of weak decays of 
hypernuclei have generated renewed interest on nonleptonic
weak interactions of hadrons.
A long standing problem is the dominance of $\di=1/2$ amplitudes
in the strangeness changing transitions.
The decays of kaons, and $\Lam$, $\Sig$ hyperons are dominated 
by the $\di=1/2$ transition but it is not clear whether this dominance
is a general property of all nonleptonic weak interactions.
In fact, the weak effective interaction which is derived
from the standard model including the perturbative QCD
corrections contains a significantly large $\di=3/2$ 
component{\cite{hweak}}.
It is therefore believed that nonperturbative QCD corrections, 
such as hadron structures and reaction mechanism are
responsible for suppression of $\di=3/2$, and/or enhancement 
of $\di=1/2$ transition amplitudes.

From this viewpoint, decays of hyperons inside nuclear medium provide
us with a unique opportunity to study new types of nonleptonic
weak interaction, that is, two- (or multi-) baryon processes, 
such as $\Lam N\to NN$, $\Sig N\to NN$, etc.
These transitions consist the main branch of hypernuclear 
weak decays because the pionic decay $\Lam\to N\pi$ is suppressed
due to the Pauli exclusion principle for the produced nucleon.

A conventional picture of the two-baryon decay process, 
$\Lam N\to NN$, is the one-pion exchange between the baryons,
where $\Lam N\pi$ vertex is induced by the weak 
interaction{\cite{ope}}.
In $\Lam N\to NN$, the relative momentum of the final state nucleon 
is about 400 MeV/c, much higher than the nuclear Fermi momentum.
The nucleon-nucleon interaction at this momentum is dominated
by the short-range repulsion due to heavy meson exchanges and/or
to quark exchanges between the nucleons.
It is therefore expected that the short-distance interactions
will contribute to the two-body weak decay as well.
Exchanges of $K$, $\rho$, $\omega$, $K^*$ mesons
and also correlated two pions in the nonmesonic 
weak decays of hypernuclei have been
studied{\cite{mesonex,Ramos}} and it is found that the kaon 
exchange is significant, while the other mesons contribute 
less{\cite{Ramos}}.

Several studies have been made on effects of quark 
substructure{\cite{CHK,MS,ITO}}.
In our recent analyses{\cite{ITO,IOMI}}, we employ an effective weak 
hamiltonian for quarks,
which takes into account one-loop perturbative QCD corrections
to the $W$ exchange diagram in the standard model{\cite{hweak}}.
It was pointed out that the $\di=1/2$ part of the hamiltonian is
enhanced during downscaling of the renormalization point 
in the renormalization group equation. 
Yet a sizable $\di=3/2$ component remains in the low energy effective 
weak hamiltonian.
We proposed to evaluate the effective hamiltonian in the six-quark wave 
functions of the two baryon systems and derived the ``direct quark''
weak transition potential for $\Lam N \to NN${\cite{ITO,IOMI}}.
Our analysis shows that the direct quark contribution
largely improves the discrepancy between the meson-exchange theory
and experimental data for the ratio of the neutron- and proton-induced
decay rates of light hypernuclei.
It is also found that the $\di=3/2$ component of the effective hamiltonian
gives a sizable contribution to $J=0$ transition amplitudes{\cite{MS}}.
Unfortunately, we cannot determine the $\di=3/2$ amplitudes 
unambiguously from
the present experimental data{\cite{Schu}}.

In this paper, we would like to show that the $\di=3/2$ two-baryon
transition amplitudes are directly related to the S-wave $\pip$ 
emission from hypernuclei.
The relation of these two amplitudes is derived from the soft-pion
theorem and is a result of the chiral structure of the weak interaction.

The $\pip$ emission from light hypernuclei, for instance, 
$^4_{\Lambda}He$, has puzzled us for a long time. 
A few experimental data suggest that the ratio of $\pip$ and 
$\pi^-$ emission from $^4_{\Lambda}He$ is about 5\%{\cite{pip-data}}.
This small ratio is expected because the free $\Lam$ decays only into 
$p\pi^-$ and $n\pi^0$.  The $\pip$ emission requires an assistance
of a proton, \ie, $\Lam+p\to n+n+\pip$.

Several microscopic mechanisms for the $\pip$ emission
have been considered in literatures{\cite{DH,CG,GT}}.
The most natural one is $\Lam\to n\pi^0$ decay followed by
$\pi^0 p \to \pip n$ charge exchange reaction (Fig.\ 1(a)).
It was evaluated for realistic hypernuclear wave functions and
found to explain only 1.3\% (according to Table II of ref.\cite{CG}) 
for the $\pip/\pim$ ratio{\cite{DH,CG}}.
Another possibility is to consider $\Sig^+ \to \pip n$ decay
after the conversion $\Lam p \to \Sig^+ n$ by the strong interaction
such as pion or kaon exchanges (Fig.\ 1(b)).
It was found, however, that the free $\Sig^+$ decay which is dominated
by $P$-wave amplitude, gives at most 0.2\% for the $\pip/\pim$ ratio.
Indeed it is clear that the $\Sig^+$ mixing and its free decay
is not the main mechanism, for experimental data suggest that
the $\pip$ emission is predominantly in the $S$-wave with the energy
less than 15 MeV.
Recently, it was proposed that a two-body process $\Sig^+ N\to n N\pip$
must be important in the $^4_{\Lambda}He$ decay{\cite{GT}}.
But its microscopic mechanism is not specified.

We here propose to apply the soft-pion theorem to the $\pip$ emission, \ie, 
in the limit of zero pion four-momentum $q$.
The soft-pion theorem for the process $\Lam p \to nn \pip(q\to 0)$
reads{\cite{JJ}}
\begin{equation}
  \lim_{q\to 0} \langle nn\pip(q)|H_W|\Lam p\rangle 
= -{i\over \sqrt{2} f_{\pi}} \langle nn|[Q_5^-, H_W]|\Lam p \rangle
\label{soft-pion}
\end{equation}
where  $Q_5^-$ is the axial charge operator, and $H_W$ is the weak
hamiltonian which describes a strangeness changing transition.
Because there is no contribution of the neutral current, $H_W$
consists only of the left-handed currents and the flavor-singlet 
right handed currents which come from the penguin-type QCD corrections.
Then the commutation relation in eq.(\ref{soft-pion}) is given in terms
of the isospin lowering operator $I_-$ as
\begin{equation}
  [Q_5^-, H_W] = -[ I_-, H_W]
\end{equation}
and therefore can be evaluated by using the isospin property of $H_W$.

As $H_W$ changes the third component of the isospin by $-1/2$ 
when it converts
$\Lam p$ ($I_3 = +1/2$) to $nn\pip$ ($I_3 = 0$),
it may contain $H_W(\di=1/2, \di_z=-1/2)$ and
$H_W(\di=3/2, \di_z=-1/2)$.  
Now it is easy to see
that $\di=1/2$ part vanishes in (2) as
\begin{eqnarray}
  [I_-,H_W(\di=1/2, \di_z=-1/2)] &=& 0 \\{}
  [I_-,H_W(\di=3/2, \di_z=-1/2)] &=& \sqrt{3} H_W(\di=3/2, 
  \di_z=-3/2) 
\end{eqnarray}
We then obtain
\begin{equation}
  \lim_{q\to 0} \langle nn\pip(q)|H_W|\Lam p\rangle 
= {i\sqrt{3} \over \sqrt{2} f_{\pi}} 
   \langle nn|H_W(\di=3/2, \di_z=-3/2)|\Lam p \rangle
   \label{eq:pipamp}
\end{equation}
Thus we conclude that the soft $\pip$ emission in the 
$\Lam$ decay in hypernuclei is caused only by the $\di=3/2$ 
component of the strangeness changing weak hamiltonian.
In other words, the $\pip$ emission from hypernuclei probes the 
$\di=3/2$ transition of $\Lam N \to NN$.

It is worth mentioning that the soft $\pip$ is emitted from the 
external lines of the $\Lam p \to pn$ weak transition (fig.\ 2)
as the vertex factor $q^{\mu}$ is to be canceled by the propagator of
the external line $\wave 1/q$ in the $q\to 0$ limit.
The reason why the $\di=1/2$ transition is not allowed can be 
understood from the fact that the $\pip$ emission from the initial 
proton, Fig.\ 2(a), and that from the final proton, Fig.\ 2(b), cancel
each other for the $\di=1/2$ hamiltonian.

Now we understand why the previous attempts to explaining
the $\pip/\pim$ ratio failed. 
Both the charge exchange process and the $\Sig^+$ decay are 
suppressed as they are induced by 
the $\di=1/2$ part of the hamiltonian.
Eq.(\ref{soft-pion}) tells us that they vanish in the soft-pion limit
or are cancelled by other diagrams.
In fact, the suppression of the free-space $S$-wave $\Sig^+\to n\pi^+$ decay
can also be explained from the soft-pion theorem.
In the same way, two-body $\Sigma^+$ decay will be suppressed for 
low-energy $\pip$,
unless it is induced by a $\di=3/2$ weak interaction.
Recently, Shmatikov proposed a new contribution, in which the weak 
$\Lam\to n\pip\pim$ vertex is followed by $\pim$ absorption by the rest 
of the nucleus{\cite{Shma}}.  There the interferences between this 
``absorption'' diagram and other diagrams of the same order (in the chiral 
perturbation theory) are neglected.  It is, however, important to 
consider all the diagrams of the same order to realize the 
suppression due to the soft-pion theorem.  A consistent study in the 
chiral perturbation theory is underway{\cite{seno}}.

In our previous study{\cite{ITO,IOMI}}, we found that the direct quark mechanism 
supplemented by the one-pion exchange transition account for the 
non-mesonic weak decays of the $S$-shell hypernuclei fairly well.
There the direct quark amplitudes contain significant $\di=3/2$
contribution especially in the $J=0$ transitions.
Now the two-baryon matrix elements in eq.(\ref{eq:pipamp}) are 
directly related to the non-mesonic decay amplitudes of hypernuclei 
by the Wigner-Eckert formula.
Indeed all the two-body matrix elements can be 
expressed in terms of three types of reduced matrix elements:
$M_0\equiv \langle NN(I=0)|| H_W(\di=1/2) || \Lam N\rangle$,
$M_1\equiv \langle NN(I=1)|| H_W(\di=1/2) || \Lam N\rangle$, and
$M_2\equiv \langle NN(I=1)|| H_W(\di=3/2) || \Lam N\rangle$.
Table 1 shows possible reduced transition amplitudes for the 
transition of the relative $S$-wave $\Lam N$ initial state 
to the $S$, $P$, $D$-wave $NN$ final states.
Now the rates of the soft $\pip$ emission from $\Lambda p$ and 
similarly those of the soft $\pim$ emission from 
$\Lambda p$ and $\Lambda n$ are given in terms of the reduced
matrix elements by 
\begin{eqnarray}
  \Gamma_{p0}^+ &\equiv& \Gamma( \Lam p (J=0) \to nn\pip) 
    =  A\,{3\over 8} (a_2^2 + b_2^2) \\
  \Gamma_{p1}^+ &\equiv& \Gamma( \Lam p (J=1) \to nn\pip) 
    =  A\,{3\over 8} f_2^2 \\
  \Gamma_{p0}^- &\equiv& \Gamma( \Lam p (J=0) \to pp\pim) 
    =  A\,{1\over 6} \left((a_1+a_2)^2 + (b_1+b_2)^2 \right) \\
  \Gamma_{p1}^- &\equiv& \Gamma( \Lam p (J=1) \to pp\pim) 
    =  A\,{1\over 6} (f_1+f_2)^2 \\
  \Gamma_{n0}^- &\equiv& \Gamma( \Lam n (J=0) \to pn\pim) 
    =  A\,{1\over 12} \left((a_1-2a_2)^2 + (b_1-2b_2)^2 \right) \\
  \Gamma_{n1}^- &\equiv& \Gamma( \Lam n (J=1) \to pn\pim) 
    =  A\,{1\over 12} (f_1-2f_2)^2 + A\,{1\over 4} (c_0^2 + d_0^2 + e_0^2)
\end{eqnarray}
where $A$ is a common kinematical factor.

The values of the amplitudes given in Table 1 are computed from those
for the non-mesonic decay of the $S$-shell hypernuclei{\cite{ITO}}.
We consider the direct quark, the one-pion exchange and the one-kaon 
exchange processes.  Among them, we assume that $\pi$ and $K$ 
exchanges contribute only to the $\di=1/2$ amplitudes, 
$a_1$, $b_1$, $f_1$, $c_0$, $d_0$,
and $e_0$, while the $\di=3/2$ parts, $a_2$, $b_2$ and $f_2$, are
purely from the direct quark mechanism.
From these values we estimate the ratios of the $\pip$ emission to the 
soft $\pim$ emission in $^4_{\Lam}He$ as
\begin{equation}
R(\pip/\hbox{soft $\pim$}) =
{\Gamma_{p0}^+ + 3 \Gamma_{p1}^+ \over
 \Gamma_{p0}^- + 3 \Gamma_{p1}^- +2 \Gamma_{n0}^-}
\approx 0.34 
\label{eq:ratio}
\end{equation}
where we have assumed that the kinetic factors such as the phase 
space integral and the final state distortions are common to all the 
channels.  If we omit the kaon exchange, the DQ + OPE gives 0.31 for 
the same ratio.

This value cannot be compared with the observed ratio
$R(\pip/\hbox{all $\pim$})$  because 
the emitted $\pim$ comes 
predominantly from the one-body ``quasi-free'' decay, $\Lam\to p\pim$.
It is therefore necessary to estimate the ratio of the soft $\pim$ and 
``quasi-free" $\pim$.
If we take the above ratio for the $\pip$ emission and the observed value
$R(\pip/\hbox{all $\pim$}) \wave$ 5\%, then we conclude that about
85\% of $\pim$ emission from $^4_{\Lam}He$ should come from the ``quasi-free''
$\Lambda\to p\pim$ decay.
Or, we may directly compare the low energy part ($E_{\pi} < 20$ MeV)
of the $\pip$ and $\pim$ spectra.
Such experimental data are most preferable.

A natural question is whether other hypernuclei emit $\pip$ as well.
Experimentally, the confirmed $\pip$ emissions are mostly attributed to
$^4_{\Lam}He$  and none to $^5_{\Lam}He$ or $^4_{\Lam}H$.
However, similar calculations as above lead to
the $\pip/\pim$ ratio, 0.35 for $^5_{\Lam}He$ and 0.77
for $^4_{\Lam}H$, respectively.
This discrepancy can be accounted for by considering the ``shell 
effect'' or the ``Pauli blocking'' in the final states, 
which is known to be important in the 
pionic decays of, for instance, $^{12}_{\Lam}C${\cite{Motoba}}.

$^5_{\Lam}He$ has the following decay modes with the corresponding Q 
values,
\begin{eqnarray}
^5_{\Lam}He &\to& ^{4}He + p + \pim +{\rm 35 MeV} \nonumber\\
     &\to& ^{4}He + n + \pi^{0} +{\rm 38 MeV} \nonumber\\
     &\to& ^{3}He + p + n + \pim +{\rm 14 MeV} \nonumber\\
     &\to& ^{3}H + n + n + \pip +{\rm 12 MeV} \nonumber
\end{eqnarray}
One sees that the Q value for the $\pip$ emission is much smaller than 
that for $^{4}He + p + \pim$ decay.  Furthermore, two neutrons emitted 
in $^{3}H + n + n + \pip$ decay are blocked by the Pauli exclusion 
principle as we have four neutrons in the final state{\cite{DH}}.  
Therefore we expect that $\pip$ decay is strongly 
suppressed in the $^5_{\Lam}He$ decay.
However, this suppression does not happen in the $^4_{\Lam}He$ decays,
\begin{eqnarray}
^{4}_{\Lam}He &\to& ^{3}He + p + \pim +{\rm 35 MeV} \nonumber\\
     &\to& ^{3}H + n  + \pip +{\rm 34 MeV} \nonumber
\end{eqnarray}
Although the outgoing $p$ and $n$ are Pauli blocked, the Q values are 
the same and the $\pip/\pim$ ratio is not affected by the Pauli 
blocking.
For $^4_{\Lam}H$, the situation is similar to $^5_{\Lam}He$, as the 
$\pip$ decay has to break $^{3}H$ nucleus as
\begin{equation}
^{4}_{\Lam}H \to 4n + \pip +{\rm 25 MeV}\nonumber
\end{equation}
Again the Q value is smaller and two of the final state neutrons are 
blocked by the Pauli principle.
On the other hand, 
\begin{equation}
^{4}_{\Lam}H \to ^{4}He + \pim +{\rm 56 MeV} \nonumber
\end{equation}
has a large Q-value and also has an extra enhancement factor due to 
the antisymmetrization.
Thus the  calculation without considering the phase space
and the Pauli blocking
corrections must seriously overestimate the $\pip/\pim$ ratios
for $^5_{\Lam}He$ and $^4_{\Lam}H$.

The vector mesons may be included in this estimate.  Most literatures
assume the octet ($\di=1/2$) dominance for the $\Lam N$-meson
couplings and therefore no contribution is expected to the $\pip$
emission, although a recent study{\cite{MS2}} suggested that the $\rho$
meson exchange may break the $\di=1/2$ rule.
It is not clear whether the direct quark mechanism and the vector meson 
exchange mechanism are related to each other.
In the present study, we have not considered isospin symmetry breaking 
in the $A=4$ hypernuclei either.

Validity of the soft-pion limit may be crucial in this study.  As far
as we consider low-energy $S$-wave pions, both its energy and momentum 
are about $100-150$ MeV and therefore are much smaller than
the chiral scale $\wave 1$ GeV.  In general, this is the energy region
where the soft-pion limit works{\cite{CA}}.
The tree level chiral perturbation theory has been questioned in a
similar pion production process (in the strong interaction) 
$ pp \to pp\pi^0${\cite{pppi0}}.
There the pion rescattering term calculated in the soft-pion limit
largely underestimates the observed $\pi^0$ production at the
threshold.
The reason for this failure seems the large off-shellness of the
exchanged pion.  Indeed, as there is a large momentum mismatch
in this process, the exchanged pion carries a significant four momentum.
It should be noted that this does not apply to our case, because we do 
not use the soft-pion approximation for the exchanged pion in our
approach.  On the contrary, we claim that the short-distance effects
represented by the direct quark diagrams are responsible for the
$\pip$ emission from hypernuclei.

In conclusion,  we have proved that the $\pip$ emission from 
hypernuclear weak decay is induced only by the $\di=3/2$ weak 
interactions in the soft-pion limit.
Therefore the $\pip$ decay due to the $\di=1/2$ processes are 
suppressed for low-energy $\pip$ and the decay will be
a clear signal of the $\di=3/2$ amplitudes
of the strangeness-changing weak interactions.
It has also been shown that the $\pip$ emission amplitudes are related
directly to those of two-body weak matrix elements.
We have evaluated the latter in the direct-quark mechanism of the 
non-mesonic decays of hypernuclei and have found the ratio,
$R(\pip/\hbox{soft $\pim$}) \approx 1/3$ 
for $^4_{\Lam}He$.
Further experimental efforts to measure $\pip$ emissions
are strongly encouraged.

\bigbreak
The author thanks Avraham Gal, Ben Gibson, Izumi Fuse, Takashi 
Inoue, and Tomoyuki Seno for useful discussions. 
He acknowledges Ben Gibson for providing their paper prior to its publication. 
He also thanks Kenji Sasaki for helping numerical computation. 
This work is supported in part by the Grant-in-Aid for scientific
research (C)(2)08640356 and Priority Areas (Strangeness Nuclear
Physics) of the Ministry of Education, Science and Culture of Japan.

\def \vol(#1,#2,#3){{{\bf {#1}} (19{#2}) {#3}}}
\def \NP(#1,#2,#3){Nucl.\ Phys.\          \vol(#1,#2,#3)}
\def \PL(#1,#2,#3){Phys.\ Lett.\          \vol(#1,#2,#3)}
\def \PRL(#1,#2,#3){Phys.\ Rev.\ Lett.\   \vol(#1,#2,#3)}
\def \PRp(#1,#2,#3){Phys.\ Rep.\          \vol(#1,#2,#3)}
\def \PR(#1,#2,#3){Phys.\ Rev.\           \vol(#1,#2,#3)}
\def \PTP(#1,#2,#3){Prog.\ Theor.\ Phys.\ \vol(#1,#2,#3)}
\def \ibid(#1,#2,#3){{\it ibid.}\         \vol(#1,#2,#3)}
\def\MO{M.~Oka} \def\etal{{\it et al.}}

\begin{table}[th]
\caption{Two-body reduced matrix elements of the weak hamiltonian for
DQ + OPE + OKE / DQ + OPE 
in units of $10^{-9}$ MeV$^{-1/2}$.}

\begin{center}
\begin{tabular}{|r|l|l|l|}
\hline
$\Lam N\to NN$ &$\langle I=0||H_W({1\over 2})||\Lam N\rangle$ &
$\langle I=1||H_W({1\over 2})||\Lam N\rangle$ &
$\langle I=1||H_W({3\over 2})||\Lam N\rangle$  \\
\hline
$^1S_0\to ^1S_0$ &$\quad-$ &$a_1\quad (3.50/ 7.47)$ &$a_2\quad  (-13.62)$ \\
$^1S_0\to ^3P_0$ &$\quad-$  &$b_1\quad  (1.00/ 7.64)$ &$b_2\quad  (-12.76)$ \\
$^3S_1\to ^3S_1$ & $c_0\quad  (-5.95/ -7.55)$ &$\quad-$ &$\quad-$  \\
$^3S_1\to ^3D_1$ & $d_0\quad  (7.65/ 13.93)$ &$\quad-$ &$\quad-$  \\
$^3S_1\to ^1P_1$ & $e_0\quad  (-5.16/ -4.13)$&$\quad-$ &$\quad-$  \\
$^3S_1\to ^3P_1$ &$\quad-$  &$f_1\quad  (11.71/ 6.30)$ &$f_2\quad (0.37)$ \\
\hline
\end{tabular}
\end{center}
\end{table}

\begin{figure}  
 \caption{Charge exchange mechanism (a) and virtual $\Sig$ excitation 
 (b) for $\pip$ emission in the $\Lam$ weak decay.}
\end{figure}
\epsfxsize=13cm \epsfbox{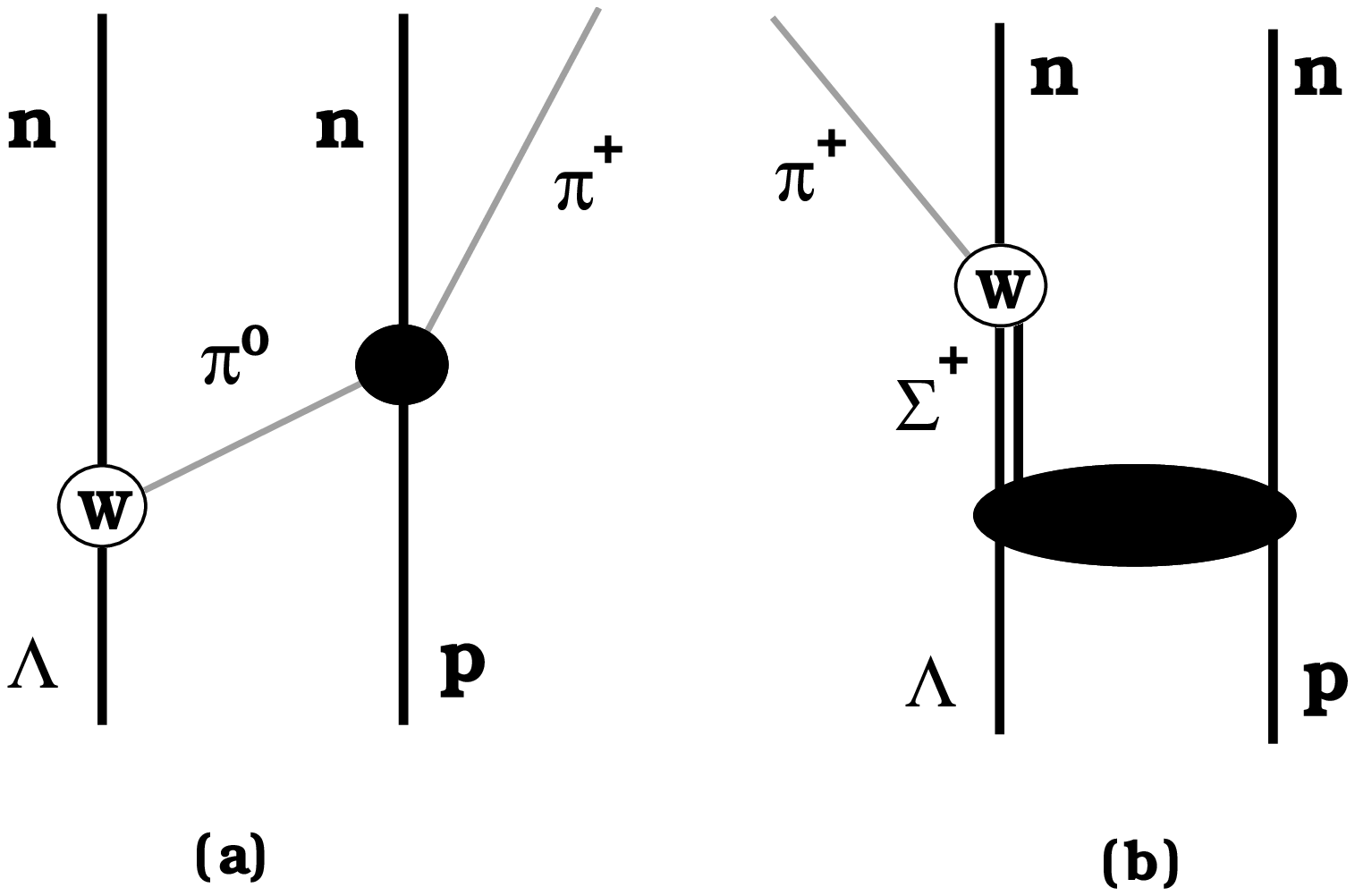}

\begin{figure}  
 \caption{Soft $\pip$ emissions in the two-body weak decay.}
\end{figure}
\epsfxsize=13cm \epsfbox{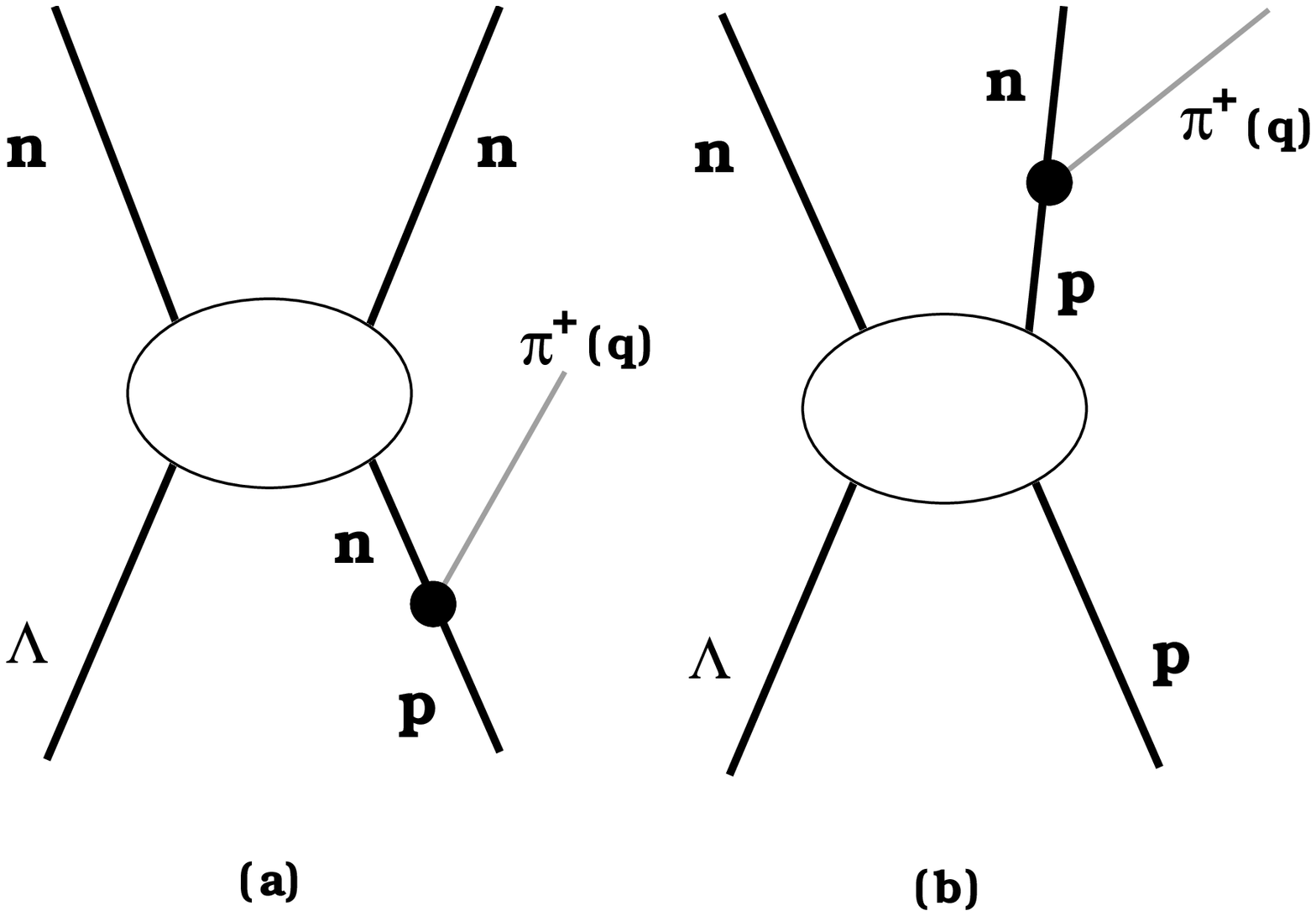}

\end{document}